\documentclass[aps,prb,twocolumn,groupedaddress,floatfix]{revtex4}
\usepackage{graphicx}
\usepackage{bm}
\begin{document}

 \title{Confinement, entropy, and single-particle dynamics of equilibrium hard-sphere mixtures
\footnote{Contribution of The National Institute of Standards and Technology, not subject to U.S. copyright}}

\author{Jeetain Mittal}
\email[]{jeetain@helix.nih.gov}
\affiliation{Laboratory of Chemical Physics, NIDDK, National Institute 
of Health, Bethesda, MD}

\author{Vincent K. Shen}
\email[]{vincent.shen@nist.gov}
\affiliation{Physical and Chemical Properties Division, National Institute 
of Standards and Technology, Gaithersburg, MD} 

\author{Jeffrey R. Errington}
\email[]{jerring@buffalo.edu}
\affiliation{Department of Chemical and Biological Engineering, University 
at Buffalo, The State University of New York, Buffalo, NY}

\author{Thomas M. Truskett}
\email[]{truskett@che.utexas.edu}
\affiliation{Department of Chemical Engineering and Institute for Theoretical Chemistry, 
The University of Texas at Austin, Austin, TX}

 \date{\today}

\begin{abstract}
We use discontinuous molecular dynamics and grand-canonical transition-matrix Monte Carlo simulations to explore how confinement between parallel hard walls modifies the relationships between packing fraction, self-diffusivity, partial molar excess entropy, and total excess entropy for binary hard-sphere mixtures. To accomplish this, we introduce an efficient algorithm to calculate partial molar excess entropies from the transition-matrix Monte Carlo simulation data. We find that the species-dependent self-diffusivities of confined fluids are very similar to those of the bulk mixture if compared at the same, appropriately defined, packing fraction up to intermediate values, but then deviate negatively from the bulk behavior at higher packing fractions. On the other hand, the relationships between self-diffusivity and partial molar excess entropy  (or total excess entropy) observed in the bulk fluid are preserved under confinement even at relatively high packing fractions and for different mixture compositions. This suggests that the partial molar excess entropy, calculable from classical density functional theories of inhomogeneous fluids, can be used to predict some of the nontrivial dynamical behaviors of fluid mixtures in confined environments. 
\end{abstract}

\maketitle

\section{Introduction}
Confinement of a fluid can substantially modify its physical properties.\cite{Drake1990,Evans1990,Gelb1999}~Although the associated changes to thermodynamics and structure can often be predicted by theory for simple fluids, estimating the implications of confinement for transport coefficients is more challenging. Since the latter remains key to understanding many systems of scientific and technological interest, the discovery of even simple heuristics could have significant impact.

In this spirit, one useful line of inquiry is to use molecular simulations to understand the behavior of simple fluid models in controlled confinement conditions. Here, we study hard-sphere (HS) fluid mixtures, both in bulk and as thin films confined between smooth hard walls. The HS model captures much of the important fundamental physics of fluids, most notably entropic packing effects associated with excluded volume interactions. While it is arguably the most basic fluid model, its historically significant role in understanding the bulk fluid-solid phase transition,\cite{alder1957,Wood1957}~ its structural similarity to simple atomic liquids,\cite{HansenMcDonald} and its ability to quantitatively describe experiments on colloidal suspensions,\cite{Pusey1986,Kegel2000,Weeks2000}~make it the standard reference system for fluids. 

Given the simplicity of the HS potential, it is perhaps surprising that numerically precise and comprehensive data for the monatomic HS fluid confined between parallel hard walls have been obtained only very recently.\cite{Mittal2007}~These data show that, for a wide range of confinement conditions, the fluid behavior is similar to that of the bulk, provided that the comparison is made using an appropriately defined density or packing fraction. Specifically, the density should be defined based on the total volume of the system as opposed to the smaller volume accessible to fluid particle centers.\cite{Mittal2006,Mittal2007}~The thermodynamic and kinetic behavior of the HS fluid confined between hard walls starts to deviate significantly from the bulk behavior only at relatively high particle densities and restrictive pore sizes.\cite{Mittal2007}~Based on comparison between the confinement shifted solid-fluid phase boundaries as calculated by Fortini {\em {et al.}},\cite{Fortini2006}~and deviations from bulk behavior in quantities like transverse diffusivity, pressure, etc. it appears that the origin of these effects is packing frustration due to the confined geometry being incommensurate with the formation of an integer number of fluid layers between the hard walls.

A second observation based on recent simulations of the monatomic HS fluid is that the relationship between excess entropy (relative to an ideal gas) and self-diffusivity is nearly identical for bulk and confined systems.~\cite{Mittal2007,Mittal2006}~In fact, excess entropy is a significantly more accurate predictor than average density for how confinement of the HS fluid affects its self-diffusivity.  A similar observation also holds for predicting the effects of confinement for the dynamics of the equilibrium Lennard-Jones and square-well fluids.~\cite{Mittal2007a}~The use of excess entropy to estimate the self-diffusivities of confined fluids was originally motivated by earlier simulations which demonstrated a strong correlation between the two quantities for a number of bulk fluids.\cite{Rosenfeld1977,Rosenfeld1999,Dzugutov1996,Dzugutov2001,Dzugutov2002,Hoyt2000,Samanta2001}~More recent simulations indicate that excess entropy (and its two-body approximation based on the pair correlation function~\cite{Nettleton1958,Mountain1971,Baranyai1989}) can also provide new insights into the connections between the static structure, thermodynamics, and dynamics of supercooled liquids.~\cite{Mittal2006a,Mittal2006b,Errington2006,Krekelberg2007}    

Can the relationships for monodisperse fluid systems discussed above be generalized for multicomponent systems?  From a practical perspective, understanding the behaviors of equilibrium mixtures will have benefits, since many experimental systems are inherently polydisperse.  There is also an urgent need for improved predictive capabilities for transport coefficients in fluid mixtures due to their relevance in industrial separation processes.\cite{Taylor2007}~Finally, understanding equilibrium mixtures will serve as a foundation for future studies of confined, supercooled liquids.  There are a number of fundamental open questions regarding these latter systems, including predicting how confinement will shift the glass transition.\cite{Alcoutlabi2005,Baschnagel2005,Mittal2004}~The reason that mixtures are important for studying supercooled, confined liquids is that their polydispersity frustrates the crystallization that otherwise rapidly occurs in monodisperse samples.  

In this paper, we use molecular dynamics and grand-canonical transition-matrix Monte Carlo simulations\cite{Errington2003,Errington2003a}~to explore whether density, excess entropy, or partial molar excess entropy can be used to forecast how confinement will affect the species-dependent self-diffusivity of hard-sphere mixtures.  One methodological outcome of our investigation is the introduction of an efficient algorithm to calculate partial molar excess entropies of mixtures from grand-canonical transition-matrix Monte Carlo simulation data.  Our main scientific finding is that previous results concerning the relationship of excess entropy and self-diffusivity of confined monodisperse systems appear to generalize well to the fluid mixtures we investigate here.  This suggests that knowledge of the behavior of bulk mixtures together with predictions for how confinement modifies their thermodynamics (from, e.g., density functional theory) can be used to accurately predict the single-particle dynamics of confined mixtures.

\section{Simulation and Theoretical Methods}

In this section, we describe the binary HS fluid investigated here and methods used to calculate its exact thermodynamic and kinetic properties.

\subsection{Model and Simulations}

We studied a binary mixture of HS particles with disparate particle diameters and masses. The ratio of the particle diameters is given by $\sigma_{1}/\sigma_{2}=1.3$ and the particle masses are proportional to their volume, i.e. $m_{1}/m_{2}=\sigma_{1}^3/\sigma_{2}^3$. We have chosen these parameter values to mimic recent experiments on binary colloidal mixtures in confinement.\cite{Nugent2007}~This choice sets the stage for future studies in which we plan to make comparisons to experimental observations in the supercooled fluid. 

We calculated the thermodynamic properties of the bulk and confined HS mixture using grand-canonical
transition-matrix Monte Carlo (GC-TMMC).\cite{Shen2005,Shen2006}~These simulations are conceptually equivalent to a series of semigrand simulations performed over a range of fluid densities stitched 
together using ghost insertion/deletion moves. The reader is referred to earlier work\cite{Shen2005,Shen2006}~for more details. The primary quantity yielded by GC-TMMC is the
particle number probability distribution. For a multicomponent system, this distribution is multidimensional and is denoted as $\Pi(\bm{N} ; \bm{\mu },V,T)$.
Mathematically,  $\Pi(\bm{N} ; \bm{\mu },V,T)$ represents the probability of observing the particle number vector $\bm{N}= (N_1,N_2,...N_i,...)$, where $N_i$ is
the number of particles of species $i$, in a system of volume $V$ at temperature $T$ and imposed set of chemical
potentials $\bm{\mu}=(\mu_1, \mu_2,...\mu_i,...)$, where $\mu_i$ is the chemical potential of species $i$. 

We simulated the bulk and confined binary HS mixture with specified activities $\xi_1 = \xi_2 = 7.8446$ and $\beta = 1.0$, where $\xi_i=$ exp($\beta\mu_i$)/$\Lambda_i^3$,  $\beta = 1/(k_BT)$ ($k_B$ is the Boltzmann constant), $T$ is temperature, and $\Lambda_i$ is the de Broglie wavelength of species $i$. We required each ($N_1,N_2$) pair to be visited by the system a minimum of two million times. We used a system volume of $V=125\sigma^3_{1}$ for the bulk fluid simulations, and we employed smooth hard walls of surface area of $49\sigma^2_{1}$ for the confined fluid simulations. In addition, to verify that system-size effects were negligible, we performed a series of simulations employing at least twice the volume. For the slit-pore geometry, we calculated density profiles at each ($N_1,N_2$) pair using one hundred bins. 

We used discontinuous molecular dynamics (DMD) simulations\cite{Rapaport}~to calculate the species-specific self-diffusivity of the binary system. For the bulk fluid, we employed a cubic simulation cell of $V$, and we applied periodic boundary conditions in all three directions.  For the confined fluid, we used a rectangular
parallelepiped simulation cell of $V = H_xH_yH_z$, where $H_z$ is the total (i.e., not the center-accessible) distance between the confining surfaces. We applied periodic boundary conditions in the $x$ and $y$ directions and we placed perfectly reflecting, smooth hard walls so that particle surfaces were trapped in the region $0 < z < H_z$. We obtained the self-diffusivity $D_i$ of the individual fluid components by fitting the long-time ($t \gg 1$) behavior of the average mean-squared displacement of the $i$th particles to the Einstein relation $\left<\Delta {\bf r}_d^2\right>_i = 2dD_it$, where $\Delta {\bf r}_d^2$ corresponds to the mean-square displacement per particle for $i$ type particles ($i=1$ or $2$) in the $d$ periodic directions ($d=2$,3 for the confined and bulk fluid, respectively). We used $N=2182$ particles for all DMD simulations. We also ran simulations with $N=1091$ and 4364 for several state points, and we found finite-size effects to be within statistical uncertainty of the simulation data. 

\subsection{Total Excess Entropy $\bm{S^{{\text {ex}}}}$ }

Knowledge of the particle number distribution $\Pi(\bm{N} ; \bm{\mu },V,T)$, which can be obtained from GC-TMMC simulation, is tantamount to knowledge of the system's free energy as a function of density and composition. One can show that the Helmholtz free energy $F$ of a system is related to the particle number distribution

\begin{equation} \label{helmholtz}
\beta F(\bm{N},V,T) = -  \ln{ \frac{\tilde{\Pi}(\bm{N};\bm{\mu},V,T)}{\tilde{\Pi}(\bm{0};\bm{\mu},V,T)} } + \beta
\sum_i \mu_i N_i
\end{equation}
\\
where $\tilde{\Pi}(\bm{N};\bm{\mu},V,T)$ is the normalized particle number distribution obtained from a GC-TMMC simulation. While $\tilde{\Pi}(\bm{N};\bm{\mu},V,T)$ is the primary quantity of interest, additional quantities can also be determined during the course of a GC-TMMC simulation.\cite{Errington2005}~For example, in this work, for a given particle number vector $\bm{N}$, we calculated the average total potential energy $U(\bm{N})$, and, in the case of the slit pore geometry, the average density profile of each species $\rho_i(\bm{r};\bm{N})$. From thermodynamics, it is known that

\begin{equation} \label{thermo_eq}
F = E - TS
\end{equation}
\\
where $S$ is the entropy and $E$ is the total energy

\begin{equation} \label{total_energy}
E = K + U
\end{equation}
\\
where $K$ is the kinetic energy and $U$ is the potential energy. It is straightforward to show that

\begin{equation} \label{real_entropy}
\frac{S(\bm{N},V,T)}{k_B}  = \ln{ \frac{\tilde{\Pi}(\bm{N};\bm{\mu},V,T)}{\tilde{\Pi}(\bm{0};\bm{\mu},V,T)} } + \beta E
(\bm{N}) - \beta \sum_i \mu_i N_i \ .
\end{equation}

To calculate the partial molar excess entropy $\bar{s}^{{\text {ex}}}_i$, it is first necessary to calculate the excess entropy
$S^{{\text {ex}}}$. In this work, $S^{{\text {ex}}}$ is the entropy difference between the system of interest and an ideal gas with the same temperature and spatial distribution of density and composition. Eq. (\ref{real_entropy}) provides the total entropy of the system in
terms of information obtained from GC-TMMC simulations and is completely general. Since this work deals with HS
systems, in what follows, the discussion is restricted to mixtures of spherical particles.

We study both homogeneous (bulk) and inhomogeneous fluid mixtures in this work. We first consider the ideal gas reference state of a homogeneous fluid. The Helmholtz free energy
$F^{ig}$ of a bulk ideal gas mixture is

\begin{equation} \label{ideal_gas}
\beta F^{ig}(\bm{N},V,T) = -\sum_i \ln{ \left( \frac{V ^{N_i}}{N_i ! \Lambda^{3N_i}_i} \right) } \ .
\end{equation}
\\
Substituting Eqs. (\ref{thermo_eq}) and (\ref{total_energy}) into Eq. (\ref{ideal_gas}), an expression for the ideal gas entropy can be obtained. It then immediately follows that the excess entropy is

\begin{eqnarray} \label{bulk_Sx}
\frac{S^{{\text {ex}}}(\bm{N},V,T)}{k_B} &=& \ln{\frac{\tilde{\Pi}(\bm{N};\bm{\xi},V,T)}{\tilde{\Pi}(\bm{0};\bm{\xi},V,T)}} \ +
\beta
U(\bm{N})  \nonumber \\
&&\sum_{i} -N_i \ln{\xi_i} + \ln{N_i!} -N_i\ln{N_i} \nonumber \\
&&+ N_i\ln{\rho_i}
\end{eqnarray}
\\
where $\rho_i$ is the bulk number density of species $i$.

Now we consider an inhomogeneous fluid mixture. In this case, the free energy of the system also depends on the density
profile of each species in the system. In other words, the free energy is a functional of the set of density profiles
$\bm{\rho} = \{ \rho_1(\bm{r}),\rho_2(\bm{r}),...\rho_i(\bm{r}),...   \}$.\cite{htdavis}~The excess entropy can be
calculated in a manner similar to the homogeneous case, where the total entropy of the system $S$ is again given by Eq.
(\ref{real_entropy}). The ideal gas Helmholtz free energy of an inhomogeneous fluid mixture is\cite{Choudhury1999,Sears2003}

\begin{equation} \label{helmholtz_inh}
\beta F^{ig}(\left[\bm{\rho}\right],\bm{N},V,T) = \int d\bm{r} \sum_i \rho_i(\bm{r}) \{ \ln{\left[ \rho_i(\bm{r})
\Lambda^3_i \right] -1 \} } \ .
\end{equation}
\\
Making use of Eqs. (\ref{thermo_eq}), (\ref{total_energy}), (\ref{real_entropy}), and (\ref{helmholtz_inh}), one can
show that the excess entropy of an inhomogeneous fluid mixture is given by an expression similar to that of the
homogeneous fluid

\begin{eqnarray} \label{inh_Sx}
\frac{S^{{\text {ex}}}(\bm{N},V,T)}{k_B} &=& \ln{\frac{\tilde{\Pi}(\bm{N};\bm{\xi},V,T)}{\tilde{\Pi}(\bm{0};\bm{\xi},V,T)}} \  + \beta U(\bm{N}) + \nonumber \\
&& \sum_{i} -N_i \ln{\xi_i} + \ln{N_i!} -N_i\ln{N_i} \nonumber \\ 
&&+ \int d\bm{r} \rho_i(\bm{r};\bm{N}) \ln{\rho_i(\bm{r};\bm{N})}
\end{eqnarray}
\\
where $\rho_i(\bm{r};\bm{N})$ is the density profile of species $i$ for a specified particle number pair
($N_1,N_2$). Eq. (\ref{inh_Sx}) is the multicomponent extension of the expression derived earlier by Mittal \textit{et al}.
\cite{Mittal2006}~In addition, notice that when $\rho_i(\bm{r})$ is uniform, the expression for the total excess
entropy of a homogeneous fluid mixture Eq. (\ref{bulk_Sx}) is recovered.

\subsection{Partial Molar Excess Entropy $\bar{s}^{{\text {ex}}}_i$}

Given the excess entropy $S^{{\text {ex}}}$ of a mixture, the partial molar excess entropy $\bar{s}^{{\text {ex}}}_i$ of component $i$ is
defined as\cite{Tester_Modell}

\begin{equation} \label{partial_molar_Sx}
\bar{s}^{{\text {ex}}}_i = \left( \frac{\partial S^{{\text {ex}}}}{\partial N_i} \right)_{T,p,N_{j[i]}} \ .
\end{equation}
\\
Notice that the derivative is taken at fixed temperature $T$, pressure $p$, and number of each species $j$ other than
$i$. However, the expressions for calculating the excess entropy using a GC-TMMC simulation, Eqs. (\ref{bulk_Sx}) and (\ref{inh_Sx}), are functions
of volume, not pressure. Therefore, it is not straightforward to take the partial derivative with respect to
$N_i$ directly while fixing the pressure. To circumvent this difficulty, we first express the total differential of $S^{{\text {ex}}}$ in
terms of $T$, $p$, and $N_i$

\begin{eqnarray} \label{dSx}
dS^{{\text {ex}}} &=& \left( \frac{\partial S^{{\text {ex}}}}{\partial T} \right)_{p,\bm{N}}dT + \left( \frac{\partial S^{{\text {ex}}}}{\partial p}
\right)_{T,\bm{N}}dp \nonumber \\ 
&&+ \sum_i \left( \frac{\partial S^{{\text {ex}}}}{\partial N_i} \right)_{T,p,N_{j[i]}}dN_i \ .
\end{eqnarray}
\\
Imposing the constraint of fixed $T$, $V$, and $N_{j[i]}$, the partial molar excess entropy can be alternatively expressed as

\begin{equation} \label{psx}
\bar{s}^{{\text {ex}}}_i = \left(  \frac{\partial S^{{\text {ex}}}}{\partial N_i}  \right)_{T,V,N_{j[i]}} - N \left( \frac{\partial
s^{{\text {ex}}}}{\partial p} \right)_{T,\bm{x}} \left( \frac{\partial p}{\partial N_i} \right)_{T,V,N_{j[i]}}
\end{equation}
\\
where $N = \sum_i N_i$, $s^{{\text {ex}}} = S^{{\text {ex}}}/N$ is the excess entropy per particle, and
$\bm{x}=(x_1,x_2,...x_i,...)$ represents the mole fraction vector for the species.

Eq. (\ref{psx}) is a general expression and provides the framework for calculating the partial molar excess
entropy using information obtained from a GC-TMMC simulation. In particular, the partial derivative taken at fixed composition suggests that a natural way to calculate
the partial molar excess entropy is along an isopleth while varying the fluid density.

To evaluate the partial derivatives in Eq. (\ref{psx}), we adopt an approach where we construct
numerically the required functions (e.g., $s^{{\text {ex}}}(p)$ at fixed $T$ and $\bm{x}$) and then fit them to polynomials,
whose derivatives can be evaluated analytically. However, from a numerical perspective, this is awkward to implement
directly from raw simulation data because $N_i$ can only take on integer values, and thus any function of $N_i$ is
discontinuous. One possible route to circumvent this difficulty involves working in terms of the mean or
ensemble-averaged analogs of the quantities in Eq. (\ref{psx}), since mean values can take on a continuous range of
values. In particular, instead of $N_i$, we use the mean particle number $\langle N_i \rangle$, which is

\begin{equation} \label{ave_N}
\langle N_i \rangle = \sum_{\bm{N}} N_i \  \tilde{\Pi}(\bm{N};\bm{\xi},V,T) \ .
\end{equation}
\\
It is likewise more convenient to work with the average excess entropy $\langle S^{ex} \rangle$,

\begin{equation} \label{ave_Sx}
\langle S^{{\text {ex}}} \rangle = \sum_{\bm{N}} \ S^{{\text {ex}}}(\bm{N},V,T) \ \tilde{\Pi}(\bm{N};\bm{\xi},V,T)
\end{equation}
\\
where $S^{{\text {ex}}}(\bm{N},V,T)$ is given by Eq. (\ref{bulk_Sx}) or (\ref{inh_Sx}). Finally, the pressure $p$, which does not
require any averaging, is simply

\begin{equation} \label{pressure}
\beta p V = - \ln{ \tilde{\Pi}(\bm{0};\bm{\xi},V,T) } \ .
\end{equation}
\\
Notice that the mean quantities and the pressure are explicit functions of the activities or chemical potentials. Given
the particle number probability distribution obtained from TMMC simulation at specified $\bm{\xi}$, the distribution
can be determined at other activities by using histogram reweighting,\cite{Ferrenberg1988}~thus allowing for the
calculation of $\langle N_i \rangle$, $\langle S^{{\text {ex}}} \rangle$, and $p$ as a function of $\bm{\xi}$, at constant volume
and temperature. Our numerical strategy, which we describe below, is to construct the required functions indicated in Eq. (\ref{psx}) by varying
$\bm{\xi}$ and then fitting them to polynomials.

\begin{figure}
\scalebox{0.45}{\includegraphics {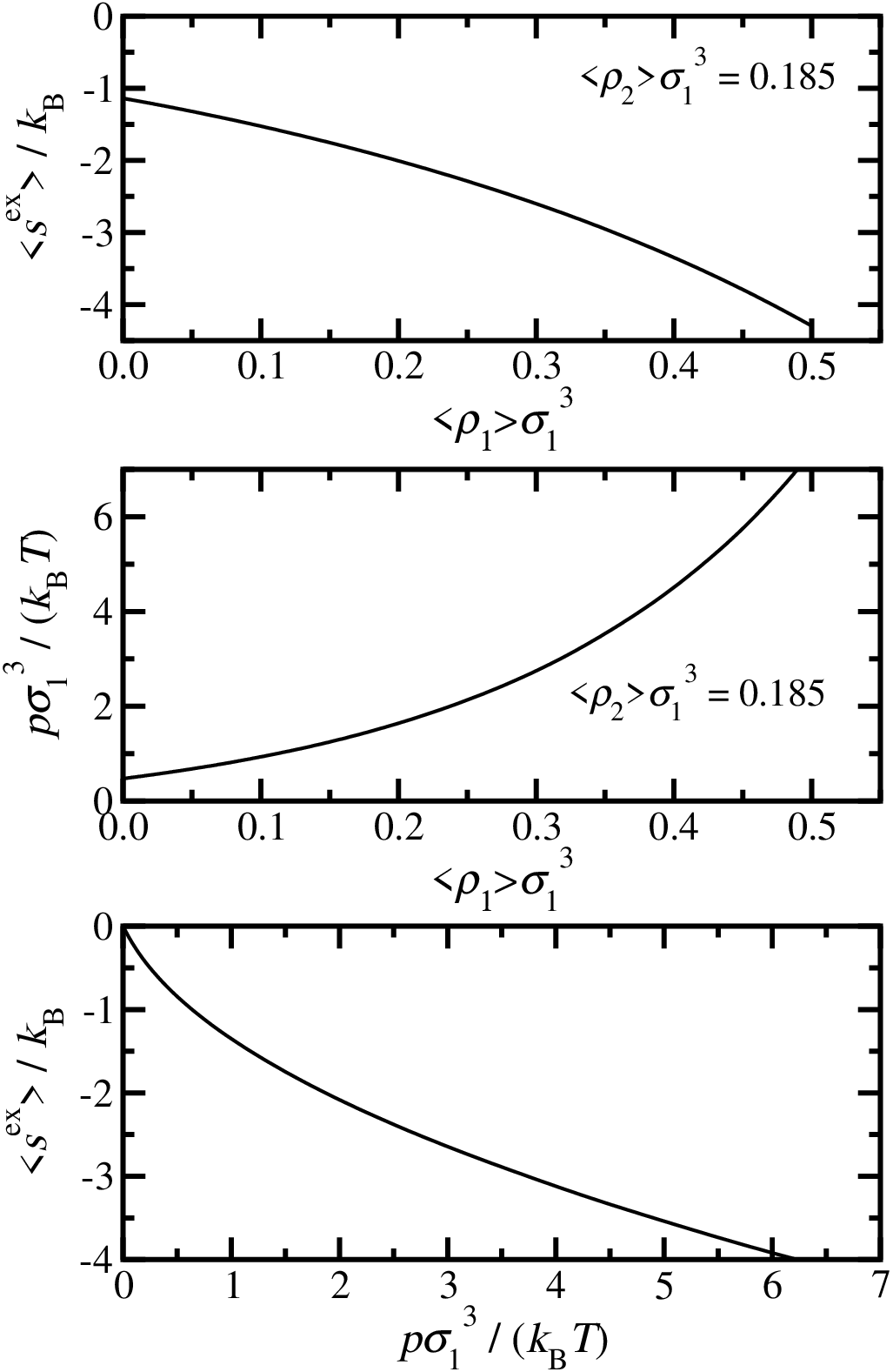}}
\caption{\label{fixedn2_fig}{ Examples of quantities needed to calculate the partial molar excess entropy. In the top and middle panel, we
  show the average excess entropy per particle and pressure, respectively, as a function of the average number
  density ($\langle \rho_i \rangle = \langle N_i \rangle/V$) of species $1$ at fixed average number density
  of species $2$. The average excess entropy per particle as a function of pressure is shown in the bottom panel
  at a fluid composition corresponding to equal volume fractions of each species. }}
\end{figure}

We first focus on the calculation of the partial derivative of the excess entropy per particle with respect to pressure
at fixed composition and temperature. For the binary HS mixture, an initial activity of one of the species,
say species $1$, is set to some arbitrary value, usually corresponding to a dilute vapor. The activity of the other
species is then adjusted such that the average composition of the fluid corresponds to the desired isopleth (within a
fractional tolerance of $1\times10^{-8}$), and the average excess entropy per particle and pressure for this activity
pair are calculated and recorded. The activity of species $1$ is then incremented and the process is repeated until
total densities ranging from vapor-like to liquid-like values are explored. This iterative process yields the average
excess entropy per particle as a function of pressure along an isopleth. In the bottom panel of Figure 1, we plot
$\langle s^{{\text {ex}}} \rangle$ versus $p$ for the bulk fluid mixture along an isopleth corresponding to equal volume fraction
of each species. Notice that the curve is smooth and continuous, allowing for the straightforward numerical
determination of its derivative.

We now focus on the calculation of the terms in Eq. (\ref{psx}) involving partial derivatives at fixed average particle
number. Conceptually, this is done by numerically constructing $\langle S^{ex} \rangle$ and $p$ as a function of
$\langle N_i \rangle$ while holding the other average particle numbers fixed. The procedure is similar to that used to
construct the $\langle s^{ex} \rangle$-$p$ curve. Consider, for illustrative purposes, the situation where $\langle
S^{ex} \rangle$ and $p$ are sought as a function of $\langle N_1 \rangle$ at fixed average number of species $2$, say
$\langle N^\circ_2 \rangle$. To do this, an initial activity $\xi_1$ is specified, and the value of $\xi_2$ is adjusted
such that $\langle N_2 \rangle = \langle N^\circ_2 \rangle$ (within a fractional tolerance of $1\times10^{-8}$). It
should be noted that simply fixing $\xi_2$ does not necessarily fix $\langle N_2 \rangle$. The
quantities $\langle S^{ex} \rangle$, $p$, and $\langle N_1 \rangle$ are then calculated and recorded for this pair of
activities, and a new value of $\xi_1$ is specified. This process is repeated until the required quantities are
obtained over the desired range of $\langle N_1 \rangle$ values. In the top and middle panels of Figure 1, examples of
these curves are shown. Again, notice that the data are smooth and continuous. From a practical point of view, one only
needs to construct the portions of these curves that coincide with the isopleth used to determine the $\langle s^{ex}
\rangle$-$p$ curve.

\begin{figure}
\scalebox{0.8}{\includegraphics{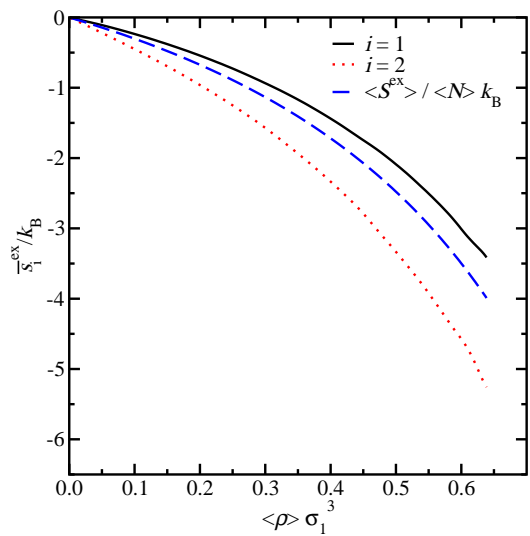}}
\caption{  \label{psx_fig}{ Partial molar excess entropy and average excess entropy per particle as a function of total fluid density
  for the bulk hard-sphere mixture at a composition corresponding to equal volume fractions of each species.}}
\end{figure}

Using the procedure outlined above, the partial molar excess entropy $\bar{s}^{{\text {ex}}}_i$ can be calculated as a function of total density
at fixed composition. In Figure 2, we plot the partial excess entropy of both species as a function of average total
density $\langle \rho\rangle = \langle N \rangle / V$ for the bulk fluid at equal volume fraction of each species.
Also plotted in the same figure is the average excess entropy per particle. As a test of thermodynamic
self-consistency, we have also verified that the average excess entropy per particle is equal to the weighted sum of
the partial molar excess entropies.

\section{Results and Discussion}
\subsection{Relationship between packing fraction and self-diffusivity}

First, we compare the self-diffusivity $D_i$ of bulk and confined fluid mixtures as a function of packing fraction $\phi$. In this subsection, the goal is to see if packing fraction alone can describe the individual component diffusivity for a confined fluid mixture. Before presenting the simulation data, we want to reiterate here our earlier proposal in recent studies that the density of confined fluids should be defined based on the total volume as opposed to the particle center accessible volume.\cite{Mittal2006,Mittal2007,Mittal2007a} When calculated in this manner, we found that packing fraction described the self-diffusivity of a number of pure simple fluids in bulk and confinement. We refer the readers to Refs.~\onlinecite{Mittal2006,Mittal2007,Mittal2007a} for further details.

\begin{figure}
\scalebox{0.9}{\includegraphics{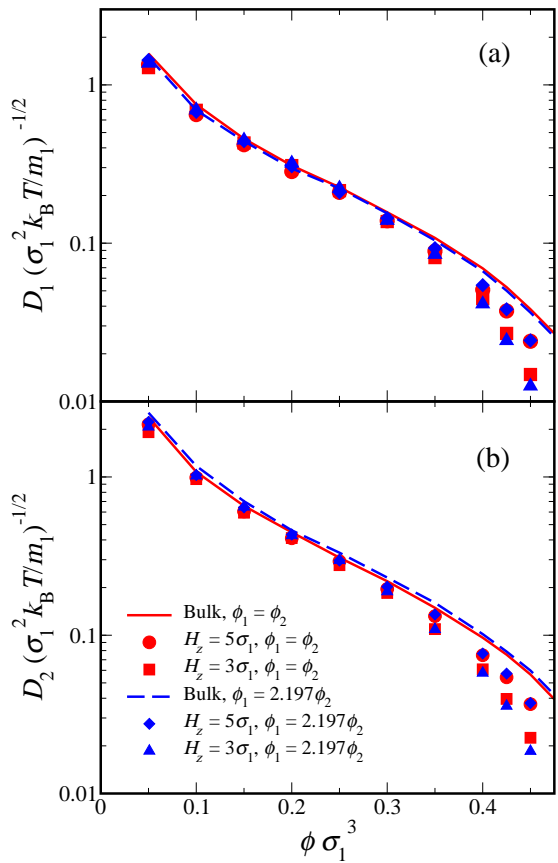}}
\caption{\label{D-Phi} Self-diffusivity versus total packing fraction for binary HS fluid in bulk and confined between hard walls.} 
\end{figure}

The predictions for the self-diffusivity $D_i$ ($i=1$, 2) versus total packing fraction $\phi$ for the binary HS mixture are presented in Fig.~\ref{D-Phi}. Here, the total packing fraction is given by $\phi = (\pi / 6) (\rho_1 \sigma_1^3 + \rho_2 \sigma_2^3)$. Data are presented for two mixture compositions, equal volume fraction ($\phi_1 = \phi_2$), and equal particle numbers ($\phi_1 = 2.197\phi_2$), i.e., equal mole fraction. The top and bottom plots correspond to the diffusivity $D_i$ of particles $i=1$ or 2 respectively and the symbols represent the confinement data as shown in the legend in Fig.~\ref{D-Phi}. It is clear from this plot that the diffusivities under confinement are very close to their bulk values up to intermediate packing fractions ($\phi <0.4$) but then deviate quantitatively at higher values of $\phi$. The diffusivities of the confined fluid can be as small as $30\%$ of their bulk values at the highest packing fraction and smallest pore-size investigated in this work ($\phi=0.45$, and $H=3\sigma_1$). Stated differently, if one uses packing fraction as a predictive tool for the diffusivity at the high packing fractions, then one will greatly overestimate the actual value. It is still remarkable that for a wide range of conditions, the data for confined fluids fall very close to the bulk curve. 

\begin{figure}
\scalebox{0.9}{\includegraphics{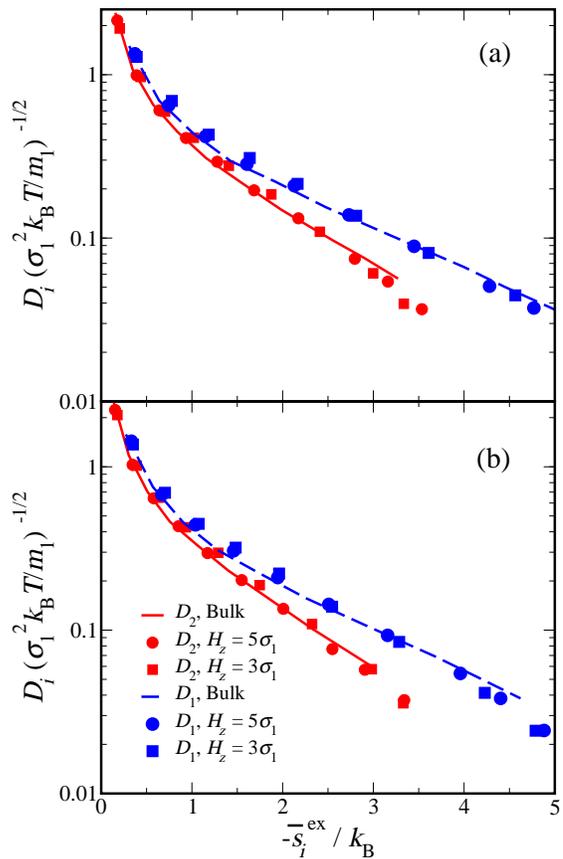}}
\caption{\label{D-exS} Self-diffusivity versus partial molar excess entropy 
 for bulk and confined binary hard-sphere fluid
between hard walls. 
(a) $\phi_1 = \phi_2$ (b) $\phi_1 = 2.197 \phi_2$. 
} 
\end{figure}

\subsection{Relationship between partial molar excess entropy and self-diffusivity}
\begin{figure}
\scalebox{0.9}{\includegraphics{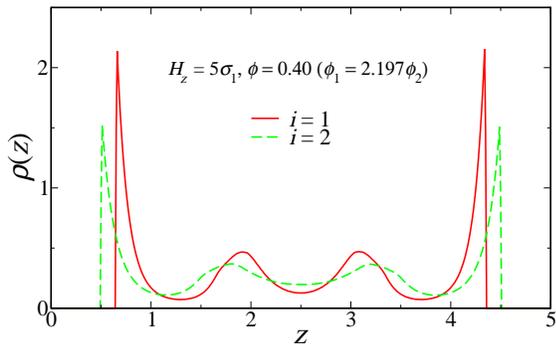}}
\caption{\label{Rhoz} Density profiles for components $1$ and $2$ for a binary hard-sphere fluid confined between hard walls.
} 
\end{figure}

Now, we investigate if a different thermodynamic quantity, i.e. partial molar excess entropy $\bar{s}^{{\text {ex}}}_i$, correlates with the confined fluid's self-diffusivity more reliably than packing fraction. Figure~\ref{D-exS} shows the self-diffusivity data for both bulk and confined fluid mixtures plotted versus $\bar{s}^{{\text {ex}}}_i$ for the same conditions as Fig.~\ref{D-Phi}. Note that the confined fluid data for a components 1 and 2 approximately collapse onto the bulk curve over the whole range. This has potentially powerful implications for predicting the diffusivity of confined mixtures. Specifically, one can use existing thermodynamic theories, such as density functional theory, for calculating the partial molar excess entropies for these mixtures in confinement and then use the bulk $D-\bar{s}^{{\text {ex}}}_i$ relationships to predict diffusivities under confinement. The fact that confinement does not significantly change the relationships between self-diffusivity and partial molar excess entropy does not mean that these two quantities remain constant when the fluid is confined. Rather, it signifies that confinement affects these quantities in a way that preserves the relationship observed between the two for the bulk mixture. In fact, we do observe confinement induced effects in our simulations. For example, changes in the self-diffusivity due to confinement can be seen in Fig.~\ref{D-Phi}. This is also accompanied by noticeable changes in fluid structure, specifically the formation of fluid layering, which can be seen in the density profiles presented in Fig.~\ref{Rhoz}.

\subsection{Relationship between total excess entropy and self-diffusivity}
\begin{figure}[b]
\scalebox{0.8}{\includegraphics{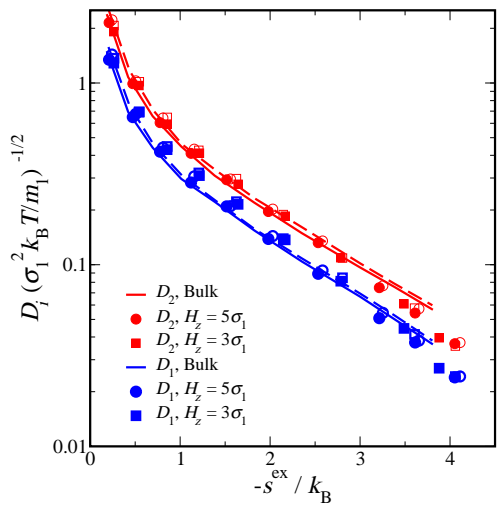}}
\caption{\label{D-TexS} Self-diffusivity versus total excess entropy 
 for binary hard-sphere fluid in bulk and confined between hard walls.
} 
\end{figure}

Here, we explore if the excess entropy itself can be used in combination with bulk fluid behavior to predict single-particle dynamics. Figure~\ref{D-TexS} shows the self-diffusivity data for (a) $\phi_1=\phi_2$ and (b) $\phi_1=2.197\phi_2$ in bulk (lines) and under confinement (symbols). For these very different compositions, one can see that the self-diffusivities collapse onto a species -specific curve independently of pore width.

\begin{figure}
\scalebox{0.7}{\includegraphics{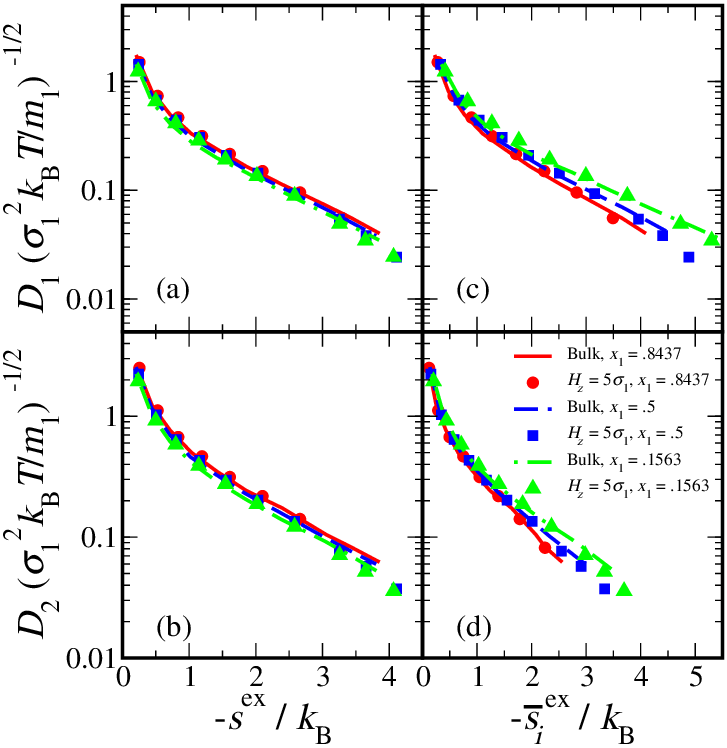}}
\caption{\label{All} Self-diffusivity versus (a,b) total excess entropy, and (c,d) partial molar excess entropy 
 for binary hard-sphere fluid in bulk and confined between hard walls. The particle mole fraction values are as shown in the legend.
} 
\end{figure}

To see if the above findings hold across the entire composition range of the mixture, we have also generated diffusivity and excess entropy data at other compositions. These additional results are given in Fig.~\ref{All}. In Figs.~\ref{All}a and \ref{All}b, we find a remarkable collapse of data when the self-diffusivity is plotted against the excess entropy. For each species, we now find that the $D_i-s^{\text {ex}}$ relationship holds, to a very good approximation, independently of pore width and composition. For completeness, the self-diffusivity is plotted against the partial excess entropy in Figs.~\ref{All}c and \ref{All}d. In this case, the data collapse independently of pore width but not composition. 

\section{Conclusions}
In this paper, we present a systematic investigation of the relationships between packing fraction, self-diffusivity, partial molar excess entropy, and total excess entropy for a binary HS mixture with components differing in their sizes and masses. To a very good approximation, the same relationship between the self-diffusivity and the packing fraction in both bulk and confined fluids is obeyed up to intermediate packing fractions. However, the deviations from this relationship at higher packing fractions and smaller pore sizes can be as high as $70\%$, indicating that the behavior of the confined fluid under such conditions differs significantly from the bulk. We find that the excess entropy and partial molar excess entropy are quantitatively more accurate predictors of single-particle dynamics (i.e., self-diffusivity) under confinement than total packing fraction. This conclusion is based upon simulation data generated over a broad range of compositions.

Another outcome of this study is the introduction of a method to calculate the partial molar excess entropy from GC-TMMC simulations and histogram reweighting. The method can be generalized to different fluid mixtures and fluid models. We are currently exploring the behavior of fluids with attractive interactions in the presence of attractive or repulsive surfaces as well as the behavior of the presently studied confined binary HS mixture in its supercooled state. In future studies, we also plan to use density functional theory to calculate the thermodynamics of mixtures, which will enable us to use the relationships between self-diffusivity and excess entropy presented here to make predictions about the single-particle dynamics of confined mixtures for a much broader range of conditions.

\section{Acknowledgments}
One of the authors (J.M.) acknowledges the financial support from a Continuing University Fellowship of The University of Texas at Austin. Two of the authors (T.M.T.) and (J.R.E.) acknowledge the financial support of the National Science Foundation under Grant Nos. CTS-0448721 and CTS-028772, respectively. The author
(T.M.T.) also acknowledges the support of the David and Lucile Packard Foundation and the Alfred P. Sloan Foundation. The Texas Advanced Computing Center (TACC) and
University at Buffalo Center for Computational Research provided computational resources for this study. A portion of this study utilized the high-performance computational capabilities of the Biowulf PC / Linux cluster at the National Institute of Health, Bethesda, MD (http://biowulf.nih.gov). This research was supported in part by the Intramural Research Program of the NIH, NIDDK.
  
\bibliography{jvjt}

\end{document}